\documentclass[conference]{IEEEtran}
\usepackage[pdftex]{graphicx}
\usepackage{paralist}
\usepackage{multirow}
\usepackage{url}

\begin{document}             
%
\title{SoS Fault Modelling at the Architectural Level in an Emergency Response Case Study}

\author{\IEEEauthorblockN{Claire Ingram, Steve Riddle, John Fitzgerald, Sakina A.H.J. Al-Lawati, Afra Alrbaiyan}
\IEEEauthorblockA{
Newcastle University, Newcastle upon Tyne, UK\\
Email: claire.ingram@ncl.ac.uk, steve.riddle@ncl.ac.uk, john.fitzgerald@ncl.ac.uk\\} }

\maketitle

\begin{abstract}
Systems of systems (SoSs) are particularly vulnerable to faults and
other threats to their dependability, but frequently inhabit domains
that demand high levels of dependability.  For this reason fault
tolerance analysis is important in SoS engineering.  The COMPASS
project has previously proposed a Fault Tolerance Architecture
Framework (FMAF), consisting of a collection of viewpoints that
support systematic reasoning about faults in an SoS at the
architectural level.  The FMAF has been demonstrated previously with
an analysis of an example fault in an emergency response SoS. In this
paper we present further examples of the FMAF's practical use, by
analysing different types of faults drawn from the same emergency
response case study.  These example faults exercise different aspects
of the FMAF, demonstrate its use in more complex fault modelling scenarios, and raise new questions for further development.
\end{abstract}

\IEEEpeerreviewmaketitle

\section{Introduction}\label{sec:intro}
Systems of systems (SoSs) face particular challenges that can increase the risk of faults, whilst typically inhabiting domains which require a high degree of dependability.  There are many examples of the types of faults that SoSs may face.  For example, communication problems can compromise the distributed constituent systems (CSs) with no warning.  CSs are likely to be different ages and make different assumptions; mismatched assumptions (e.g., architectural mismatch~\cite{Garlan&95}) are a real risk.  CSs may be independently managed, may be unaware of the SoS or reluctant to participate in it, and may evolve without considering the needs of the SoS or providing advance notice, so there is a high risk of unanticipated failures in SoS functionality.  Error propagation may be complex, governance difficult to track, and recovery strategies may not be clear.  This is particularly a problem in cases where the failure of one CS may result in an SoS-level fault which must be detected by another CS,  or which could be detected by one of several separate CSs, and in cases where CSs which may be required to implement extra (possibly costly) recovery activities.  If CSs have separate ownership and motivations, there is often a need to consider the options in advance and agree which CSs are responsible for detecting errors or for implementing and executing recovery strategies.

For these reasons, we argue that an architectural approach to fault modelling is
particularly useful for the SoS reliability engineer.  The
COMPASS\footnote{http://www.compass-research.eu/} project has
developed a Fault Modelling Architectural Framework (FMAF) that
provides a systematic approach to capturing fault tolerance aspects of
SoSs.  The COMPASS FMAF has been described in previous
publications~\cite{Andrews2013, Andrews&13d, Andrews&14} and its
practical use demonstrated with the modelling of a single fault drawn
from an emergency response SoS.  In this paper we present an extension
of previous work, modelling further faults extracted from the same
case study.  The additional faults exercise new aspects of the FMAF,
such as the inclusion of faults which may be detected by multiple 
constituent systems.

The rest of this paper is laid out as follows:
Section~\ref{sec:related} briefly summarises some related work and
Section~\ref{sec:fmaf} introduces the FMAF.
Section~\ref{sec:casestudy} introduces our case study and previous
work.  Sections~\ref{sec:erufails} and~\ref{sec:wronglocation} study
two separate faults using the FMAF approach.  Finally
Section~\ref{sec:concs} presents our conclusions.  Work presented here
forms part of the COMPASS project, building further on work presented
in~\cite{Andrews&13d}.

\section{Related work}\label{sec:related}
Fault tolerant architectures have been widely studied in the
literature (e.g., see~\cite{Rugina&08}).  Architectural approaches are
useful for modelling fault tolerance within an SoS.  Some previous
researchers have adopted an architectural approach, demonstrating 
feasibility.  For example,~\cite{Bernardi&07}
and~\cite{Bondavalli&01} use UML to model erroneous behaviour in
embedded systems, whilst~\cite{David&10} use SysML for analysis of
dependable complex physical systems. SysML is also used
in~\cite{Petin&10} for verifying safety requirements in embedded,
safety-critical control systems.

\section{Fault Modelling Architectural Framework}\label{sec:fmaf}
The COMPASS Fault Modelling Architectural Framework (FMAF) encompasses
a series of viewpoints to aid the SoS reliability engineer.  In particular it supports: 
\begin{itemize}
\item Definition of faults, errors and failures in an SoS
\item Identification of and reasoning about causal chains
\item Definition and
identification of the boundaries of CSs
\item Definition of erroneous states and recovery
scenarios.  
\end{itemize}
Table~\ref{table:fmaf} (taken from~\cite{Andrews&14})
presents the viewpoints of the FMAF, and briefly summarises their
purpose.

\begin{table*}[!t]
\begin{center}
\caption{Informal description of the FMAF viewpoints (taken from~\cite{Andrews&14})}
\label{table:fmaf}
\renewcommand{\arraystretch}{1.15}
\begin{tabular}{|c|l|l|}
 \hline
& \textbf{Name} & \textbf{Description}\\
 \hline
\multirow{5}{*}{\rotatebox[origin=c]{90}{\textbf{Structural}}}

&  Fault/Error/Failure Definition  & Define faults, errors and failures of the SoS. Faults, errors or failures may 
                                   be generalised into abstract categories.\\
\cline{2-3}
& Threats Chain View              & Identifies progression of faults through errors to failures and relationships 
                                   between the dependability threats and the constituents.  \\
\cline{2-3}
&  Fault Tolerance Structure      & Shows the composition of the SoS with the required redundancy to tolerate a given fault. \\
	
\cline{2-3}
&  Fault Tolerance Connections    & Shows connections and interfaces between constituents of the SoS with the required 
                                   redundancy to tolerate a given fault.  \\
&	                         & Includes all the constituents identified in the respective \emph{Fault Tolerance Structure View}.\\
 \hline
 \hline
\multirow{6}{*}{\rotatebox[origin=c]{90}{\textbf{Behavioural}}}
& Erroneous/Recovery Processes     & Identifies the processes of the SoS, including erroneous behaviour and any required recovery processes.\\
\cline{2-3}
& Erroneous/Recovery Scenarios     & Models behaviour in the presence of errors (with and without recovery) as scenarios. 
                                    Shows erroneous behaviour propagation  \\
&	                          & and recovery procedure triggers.\\
\cline{2-3}
&                                    Fault Activation & Defines the behaviour within an SoS process and identifies when 
                                    faults may be activated,  what happens after activation and \\
&	 		          & where in the process the error may be detected.\\
\cline{2-3}
& Recovery                         & Defines the behaviour of the recovery procedures that are triggered once an error has been detected.\\
 \hline

\end{tabular}
\end{center}
\end{table*}
When discussing fault tolerance in SoSs, we adapt a well-known dependability taxonomy which was initially provided
by~\cite{Avizienis04}.  Under this taxonomy, a \emph{failure} is a deviation from expected
service, visible outside a given system.  At the SoS level, the failure is a deviation from SoS-level service.  An \emph{error} is part of
the SoS state that can lead to a failure.  A \emph{fault} is the cause
of an error.  We note that, within an SoS, a failure at the level of a
single CS becomes a fault at the level of the SoS.  A set of rules for ensuring consistency between the viewpoints is described in~\cite{Andrews&13d}  

\section{Case Study and Previous Work}\label{sec:casestudy}
Our case study is an emergency response SoS, supplied by the Italian
company Insiel.  Insiel supports an SoS in northern Italy incorporating
separate emergency services (such as fire departments and ambulance
services).  The services are separately managed and funded.  The SoS
provides a single emergency point of contact for the public, and
delivers the appropriate aid within a target time frame. The case study has been described in previous
publications~\cite{Andrews2013, Andrews&14, Andrews&13d} that illustrate
the use of the FMAF using a single potential fault drawn from the case study (the failure of the
radio system used for communications).  

In our analysis, we concentrate on the services within the emergency response SoS that supply medical
aid.  For the purposes of the fault tolerance analysis presented here, the constituent systems of the emergency response SoS include:
\begin{itemize}
\item A system of distributed, mobile Emergency Response Units (ERUs), comprising one or
more medical staff and a driver.  
\item A radio system employed for communications
\item A mobile phone system employed as a backup for the radio
\item A central call centre which consists of expert operators and software (`CUS') to provide a workflow
\end{itemize}

In~\cite{Andrews&13d} the nominal behaviour of the SoS and initial
recovery processes for one fault (failure of the radio system) are presented in SysML.  The simplified nominal behaviour of the CSs described above can be briefly summarised as follows:
\begin{itemize}
\item Call centre receives and processes emergency calls using the CUS
\item Call centre operator finds an appropriate ERU (which may be idle or already on a
  lower priority assignment) and dispatches it to the new emergency  
\item The ERU travels to the target, dispenses aid and sends updates on current status over the radio to the call centre 
\item The call centre workflow software (the `CUS') tracks specified metrics
  e.g. time to arrive
\item The ERU staff decide on the next action: either further aid is not
  necessary and the ERU returns to idle state; or the casualty
  is transported to hospital for further treatment
\end{itemize}

\begin{table}[!t]
\begin{center}
\caption{Emergency Response SoS faults of interest  (taken from~\cite{Andrews2013})}
\label{table:faults}
\begin{tabular}{|l|l|l|l|}
	\hline
	 \textbf{Fault} & \textbf{Description}  & \textbf{Error} 		& \textbf{Recovery}\\
			&			& \textbf{Detected by}	        & \\
	\hline
	\hline
	  Fault 1   & Complete  				& ERU and/    & The ERU driver uses his/her mobile \\
			& failure of 				& or call 	 	& phone. If there is no mobile phone \\
			& the radio   				& centre       	& coverage, the personnel uses \\
			& system 				& 		 	& landline phones (e.g. the phone \\
			&						&			& of the patient)  \\
	 \hline
	  Fault 2 	& An ERU  				& ERU 		 & ERU with patient: a new ERU \\
			& breaks down 			& crew or 	& without medical personnel  is sent, \\
			& or crashes				& call centre 	& to retrieve the patient and the ERU \\
			&						&			& crew. ERU driver waits for recovery \\
			&						&			& ERU without patient: a new ERU or \\
			&						&			& a car is sent to retrieve ERU crew \\
			&						&			& (except driver, as he waits for  \\
			&						&			& assistance) \\
	   \hline
	
	  Fault 3 	& An operator				& Call centre 	& Re-direction of the ERU to the \\
			& sends an  				& operator, 	& correct target location (assuming \\
			& ERU to the  	    		& caller or 	& that this can be resolved from the \\
			& wrong 		 			& ERU 		& information available), or select \\
			& location					&			& closer ERU\\
	\hline

\end{tabular}
\end{center}
\end{table}
The nominal behaviour of the system described here is
our starting point for describing the emergency response SoS.  Three example faults for this SoS were proposed in~\cite{Andrews2013} (also shown in Table~\ref{table:faults}).  Fault 1 has been modelled using the FMAF viewpoints
previously~\cite{Andrews2013}; in this paper we present FMAF models and analysis of Faults 2 and 3 for the first time.  We consider Fault 3 first, in Section~\ref{sec:wronglocation}, presenting a selection of FMAF viewpoints to model the SoS behaviour.  There are several possibilities for presenting Fault Activation Views; we present one approach in our analysis of Fault 3, and in analysis of Fault 2 we present a second approach, for comparison.  The alternative method adopted in Fault 2 is described in Section~\ref{sec:erufails}.  

\section{Fault 3: ERU in Incorrect Location}\label{sec:wronglocation}
Fault 3 arises when `an operator sends an ERU to wrong
location'~\cite{Andrews2013}. 

\subsection{Threats Chain Views}\label{sec:3-FPVs}
The FMAF Threats Chain View (TCV) illustrates where in the SoS: 
\begin{itemize}
\item the initial \emph{fault} is located;
\item the \emph{error} can be detected; and
\item the \emph{failure} can be observed (this should be outside the
  SoS, as failures are observable at the external boundary of the
  system/SoS)
\end{itemize}

The TCV requires us to address the question \emph{which CS is responsible for activating the initial fault?}.  On considering all possibilities, we find that the same outcome (the ERU receives incorrect
information from the operator) may have several
quite different sources: the caller; the call centre operator; the
radio system; or the ERU crew.  Dependent on this, the fault is detectable in different ways by different
CSs, so
we partition the fault for separate analysis as follows:
\begin{enumerate}
\item \textbf{Fault 3.1:} the caller provides poor quality location
  information
\item \textbf{Fault 3.2:} the call centre operator receives good
  information from the caller, but makes a mistake and provides
  incorrect location when transcribing this, or when transmitting it
  to the ERU over phone or radio
\item \textbf{Fault 3.3:} the radio quality is poor and the ERU
  mis-hears the location
\item \textbf{Fault 3.4:} the call centre and radio provide good
  information, but the ERU crew makes a mistake when transcribing or recording the
  target destination
\end{enumerate}
We do not have space here for a full analysis of all four faults; we
concentrate on the scenarios presented by 3.1 and 3.2 as representative samples.  

\begin{figure}[!t]
	\centering
        \includegraphics[width=1\columnwidth]{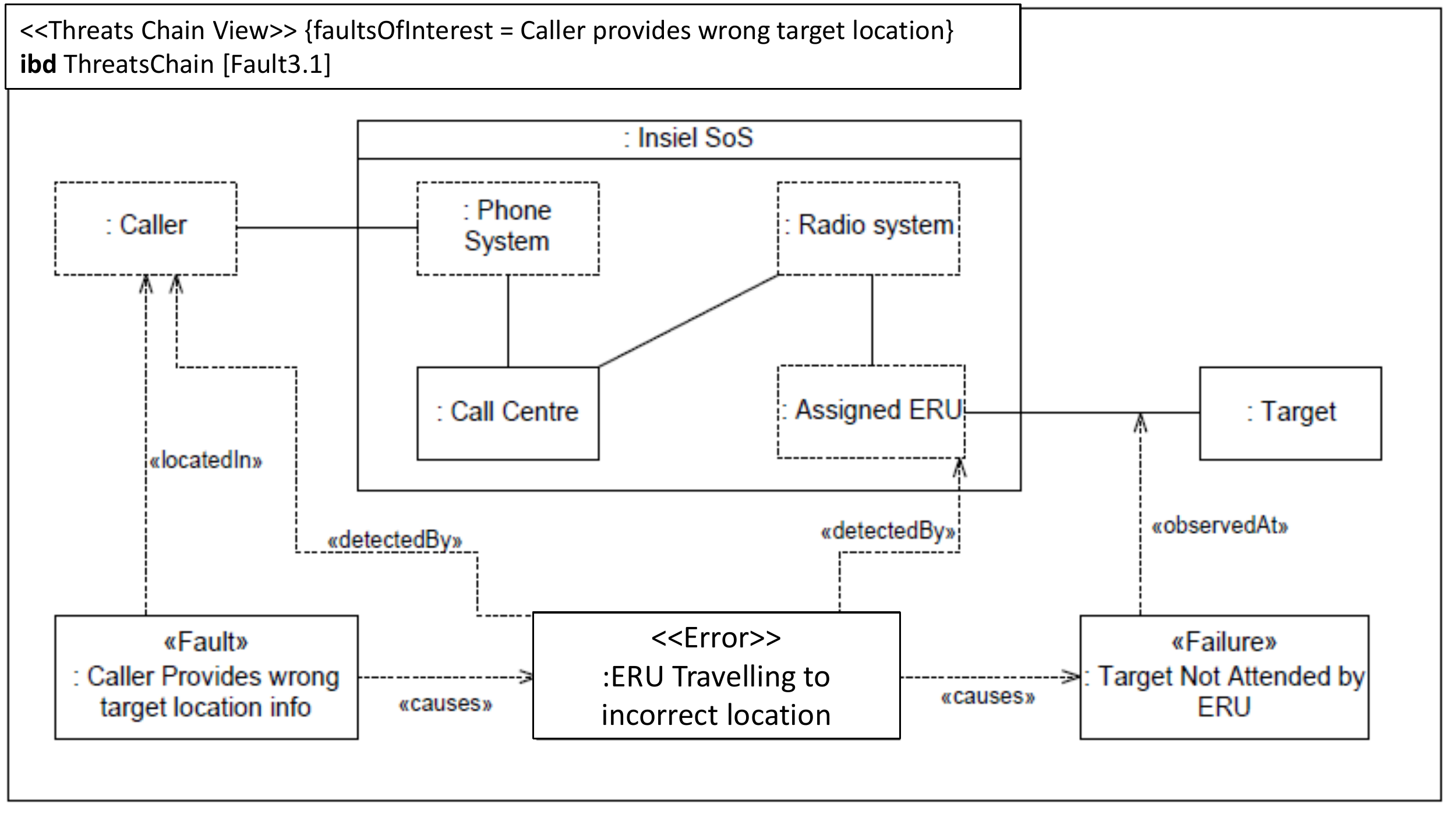}
	\caption{Threats Chain View of Fault 3.1}
	\label{fig:fault3b_incorrect}
\end{figure}

\begin{figure}[!t]
	\centering
        \includegraphics[width=1\columnwidth]{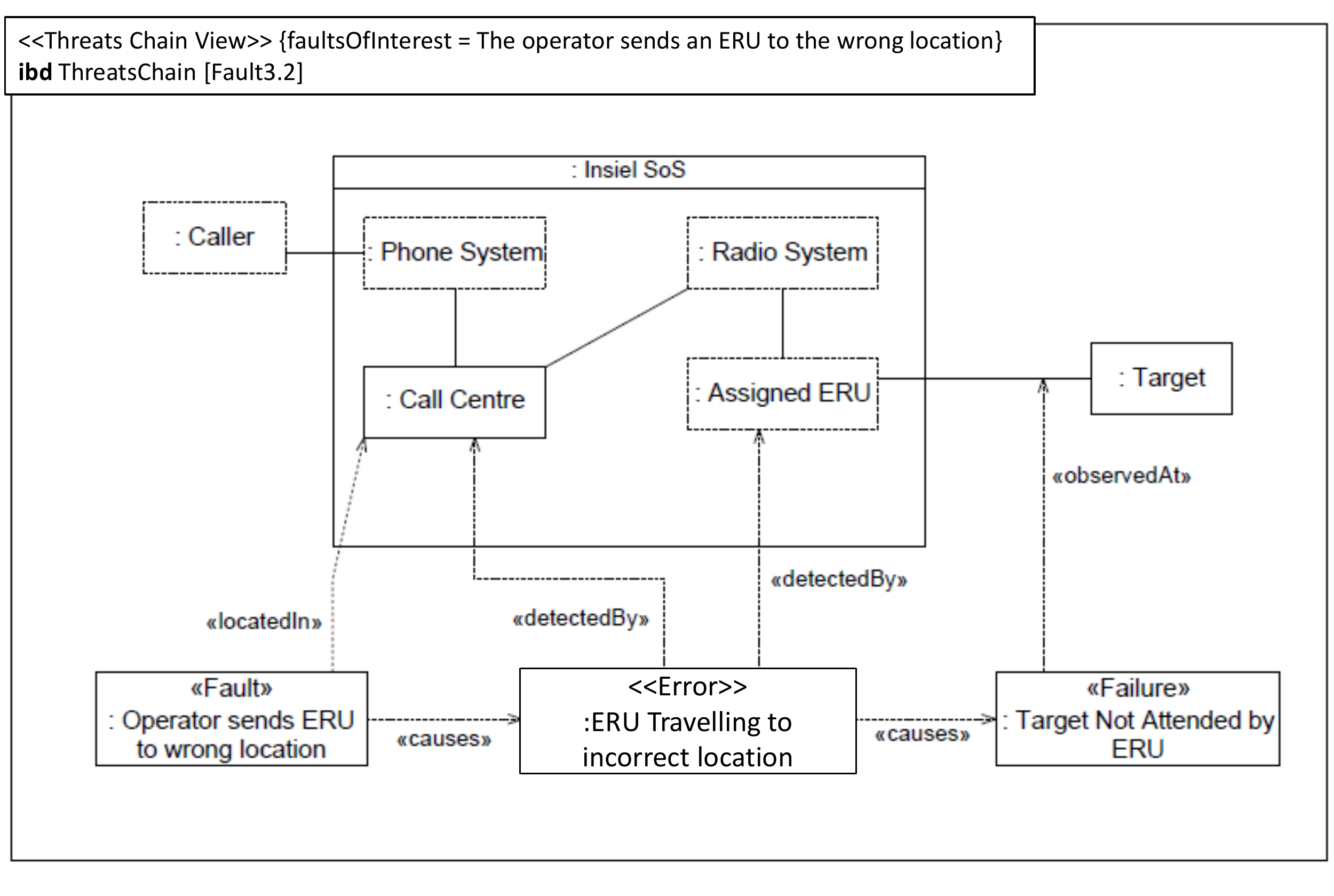}
	\caption{Threats Chain View of Fault 3.2}
	\label{fig:fault3a_correct}
\end{figure}

Examples of TCVs are presented in Figure~\ref{fig:fault3b_incorrect} (for Fault 3.1) and Figure~\ref{fig:fault3a_correct} (for Fault 3.2).  According
to our dependability taxonomy, a fault should be introduced by a CS,
and visible to external entities only at the boundary of the SoS (as a
failure).  In Figure~\ref{fig:fault3b_incorrect}, however, the caller both introduces the fault
(incorrect location information), and also is responsible for
detecting it, which is not correct according to our dependability
taxonomy.  Fault 3.1 is not in fact a `fault':
the SoS is behaving as expected, and the poor outcome is a direct
result of poor information received from the environment (we regard
the caller as part of the environment, not a CS).  The TCV is not a good vehicle for modelling propagation of
incorrect input data; the concept of a an internal fault arising and
propagating does not apply and we regard Figure~\ref{fig:fault3b_incorrect} as an incorrect usage of the view.  However, we do desire the emergency SoS
to cope well with the inevitable eventuality that some information
received from the environment will be of poor quality; this needs to be considered in the design of nominal and recovery behaviour, so although the TCV is not appropriate for Fault 3.1, other FMAF approaches are still useful.  

Figure~\ref{fig:fault3a_correct} shows that Fault 3.2 is activated in
the Call Centre constituent system.  In this situation the fault introduced by the call system becomes an error state at the level of the SoS, which can be detected by either the ERU or the call centre itself.  Finally, if no detection and recovery take place, the error propagates to a failure visible at the external boundary of the SoS as a failure to attend the target casualty.  

\begin{figure}[!t]
	\centering
        \includegraphics[width=1\columnwidth]{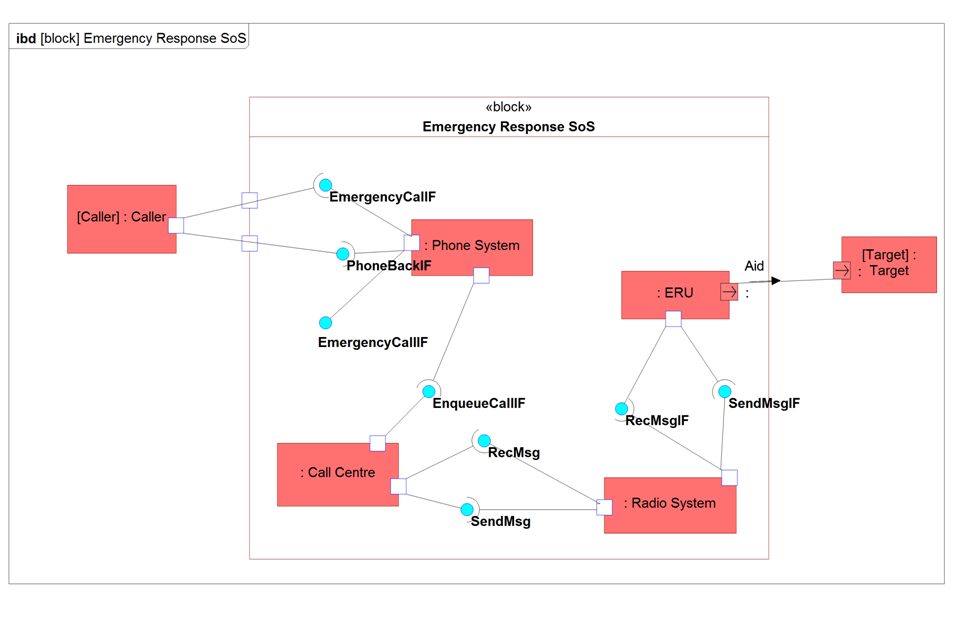}
	\caption{Fault Tolerance Connections View (FTCV) for Fault 3.1}
	\label{fig:fault3a_FTCV}
\end{figure}
\subsection{Fault Tolerance Connections view}\label{sec:3-FTCs}
In~\cite{Andrews&13d} a nominal `connections view' was presented for the
SoS, in which:
\begin{itemize}
\item A Phone System provides a connection to the external environment
  (the Caller) and the Call Centre, permitting the two to communicate.
\item A Radio System supplies interfaces to, and requires interfaces
  from, the ERU and the Call Centre, permitting these two CSs to
  communicate.
\item The ERU provides a connection to the external environment in the
  form of the target casualty that the ERU attends.
\end{itemize}
The caller and the target are modelled separately because they may not
be in the same location.  The Phone System and Radio System are
modelled explicitly because they are key systems upon which the SoS
depends, but are associated with some known risks (e.g., radio
reception is not possible in some mountainous or remote areas).  Their
inclusion permits us to model our reliance on them explicitly.  The
Phone System provides two possible connections; one for the use of
external entities, and a separate one for the use of CSs inside the
SoS.

The FMAF Fault Tolerance Connections View (FTCV) shows the internal
and external SoS connections which may be necessary for fault
tolerance.  We present an FPCV in Figure~\ref{fig:fault3a_FTCV},
depicting the same SoS internal and external connections described
above and in~\cite{Andrews&13d}, but adapted to cope with Fault 3.1.
In this scenario, the original caller provides poor quality location
information.  In some cases, after acquiring better information, the caller wishes to recontact the
call centre.  This may require consideration to be paid to connections between the SoS and the original caller, in order to 
to allow the caller to phone back with new details.  For example, the caller
may be provided with details of the specific dispatch office that is
dealing with the emergency, or perhaps the Phone System can recognise
an incoming call from a number already associated with an ongoing
emergency, and pre-populate the operator's screen with known details.   

Alternatively, in some situations it is possible that the caller does not provide quality location information and does not recontact the call centre (e.g., they may hang up or be cut off before the operator can reconfirm address details or may be too distracted to provide a quality address) and although it may be possible to ascertain the location of many callers from their phone number this is not always the case (e.g., the caller may not be in the same location as the target, or may be in a moving vehicle).  In a situation like this, responsibility for detecting the error falls upon the ERU, which will realise the problem on
arrival at the incorrect location, and then will contact the call centre for advice.  After
confirming the location of the ERU matches the location that was provided by the caller, the call centre operator needs further information about the correctness of the information they provided.  Sometimes the operator may need to re-contact the caller to confirm the correct location.
For this reason, the FTCV model for Fault 3.1 shows an additional
connection that allows the call centre operator to phone the original
caller back.  Other variations of Fault 3 may also require modified or additional connections within the SoS.  For example, Figure~\ref{fig:fault3a_FTCV} does not include the Mobile Phone System which can be used as a backup in the event of a Radio System failure, but the connections with the Mobile Phone system become important when modelling Fault 3.3 (the radio quality is too poor to hear the location clearly).  An FTCV for Fault 3.3. may depict the ERU crew making use of the Mobile Phone system to connect to the EmergencyCallIF provided by the Phone System.

\subsection{Fault Activation Views}\label{sec:3-FAVs}
\begin{figure}[!t]
	\centering
        \includegraphics[width=1\columnwidth]{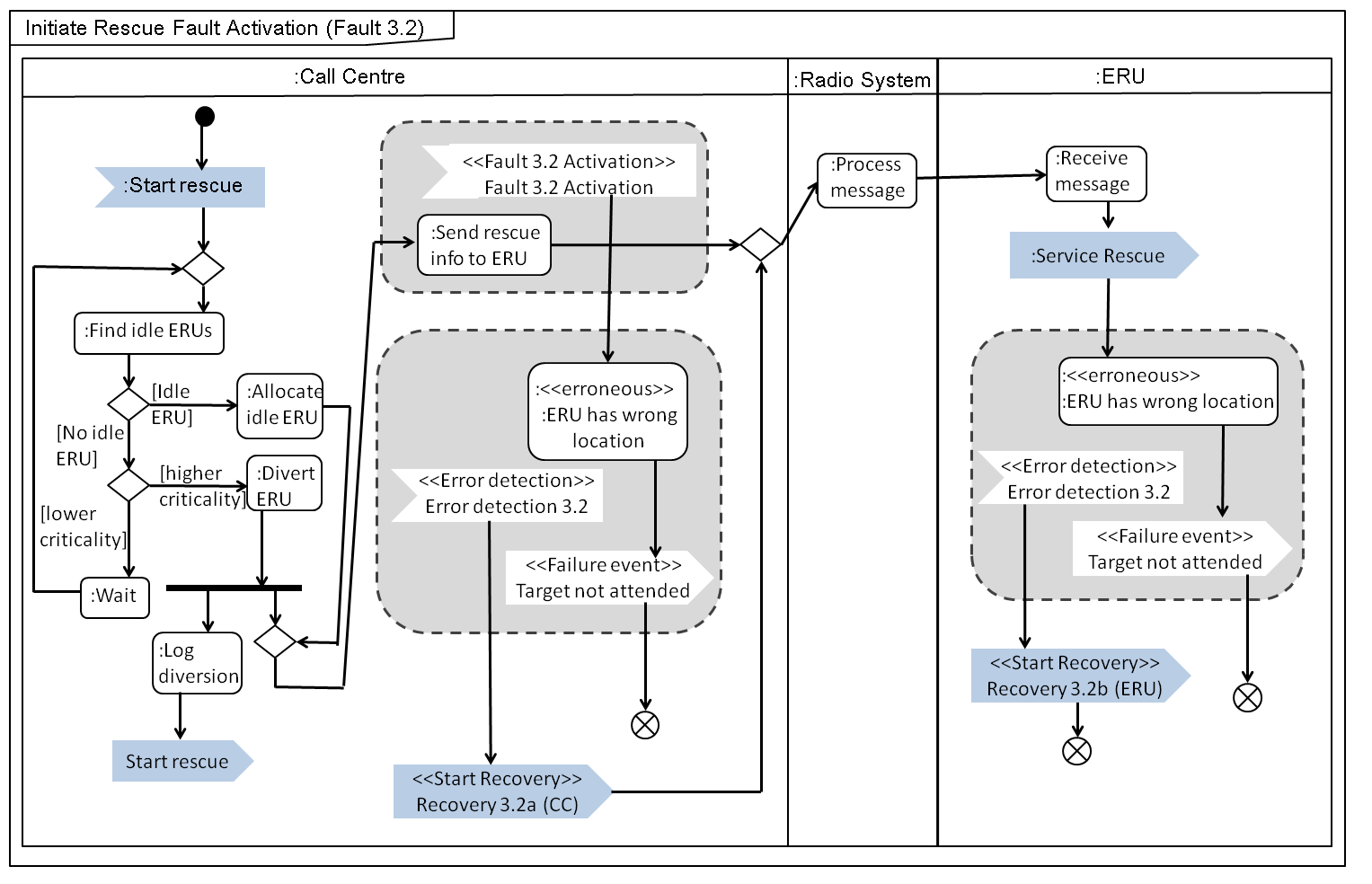}
	\caption{Fault Activation View for Fault 3.2}
	\label{fig:3b_FAV}
\end{figure}
Fault Activation Views (FAVs) are behaviour models which identify where, within the nominal behaviour, a fault can arise and which CS(s) will detect the fault.  Figure~\ref{fig:3b_FAV} presents a Fault Activation View (FAV) for Fault 3.2 (the call centre operator introduces the fault by providing incorrect details to the ERU).  The fault's activation, detection and recovery are represented using interruptible regions.  There are three interruptible regions in this FAV; one to represent the activation of the fault (within the call centre), and two representing opportunities to detect the resulting erroneous state and initiate recovery.  Two CSs are capable of detecting the error: the operator may detect this themselves (e.g., by checking recordings of the call) and recontacting the ERU; or alternatively the ERU may detect the erroneous state after arriving at an incorrect address.  Each detection event prompts the beginning of the recovery process; for example, in Figure~\ref{fig:3b_FAV}, the error detection leads to the process `Start Recovery 3.2a (CC)' or `Start Recovery 3.2b (ERU)'.  These recovery processes differ very slightly; for example, if the ERU detects the fault, then the ERU will need to contact the call centre for advice.  However, both recovery processes involve the operator clarifying the genuine emergency location (e.g., via recordings of the original call) and transmitting it to the ERU, or selecting an alternative ERU if it is closer (omitted here).

\subsection{Erroneous/Recovery Processes and Scenarios}\label{sec:3-ERSP}
Analysis of Fault 3 has identified some new recovery procedures to be
designed and enacted when specific faults are uncovered.  For example,
we can see that procedures need to be designed for: the call centre operator, for situations when more information about location is required or where the original caller phones back with extra location
details; and for the ERU, when arriving at an incorrect location.  We model these using
the FMAF Erroneous/Recovery Processes and Erroneous/Recovery Scenarios
(omitted due to space limitations).  

\section{Fault 2: ERU Breaks Down or Crashes}\label{sec:erufails}
Fault 2 arises when an ERU has broken down or crashed.    

\subsection{Fault Activation View and Recovery View}
\begin{figure}[!t]
	\centering
        \includegraphics[width=1\columnwidth]{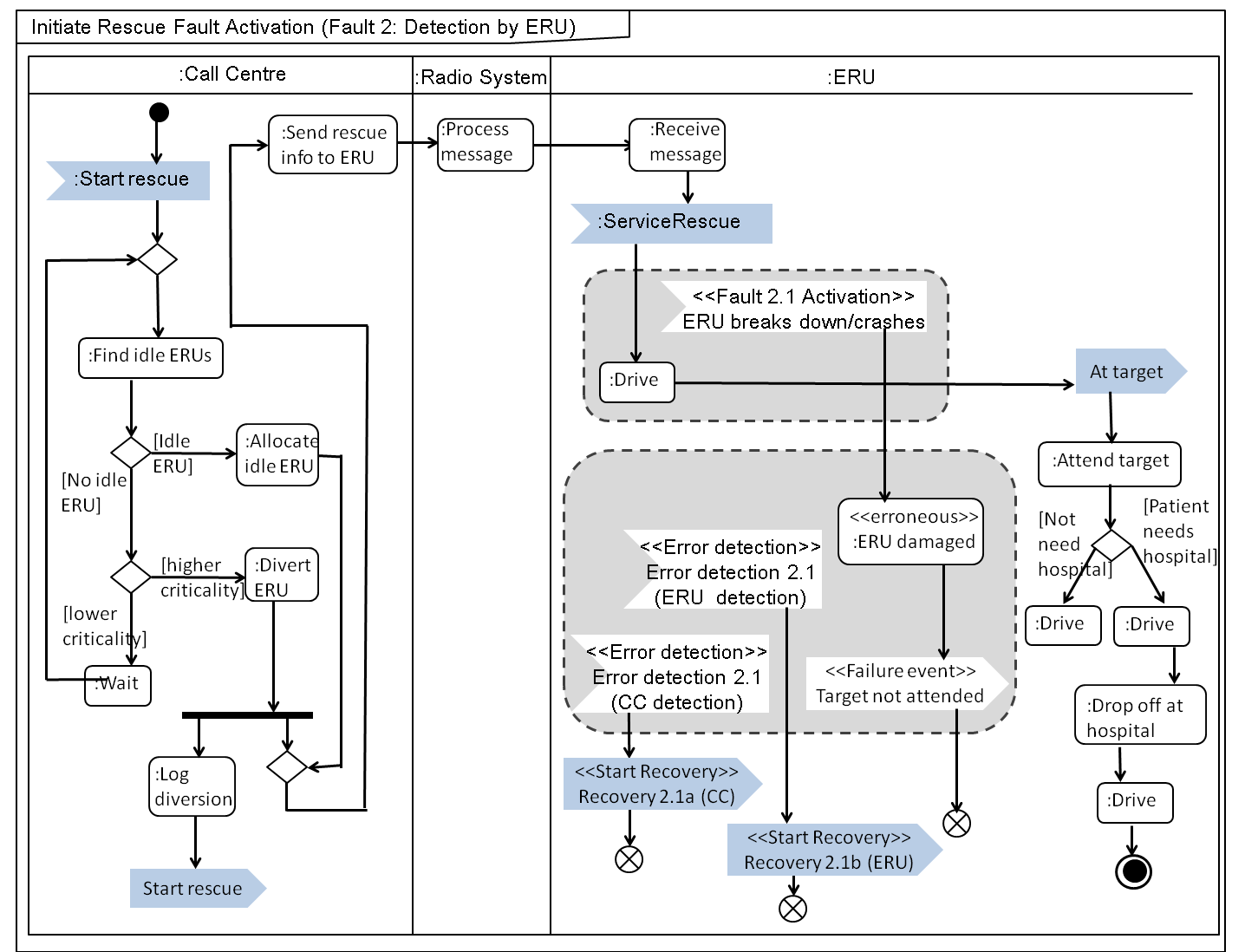}
	\caption{Fault Activation View of Fault 2}
	\label{fig:fault2-FAV}
\end{figure}
An FAV for Fault 2 is presented in Figure~\ref{fig:fault2-FAV}. In trying to answer the question of \emph{which activity is interrupted}, we find that the fault may arise in one of several different stages of ERU behaviour.  The possibilities are:
\begin{itemize}
\item \textbf{Fault 2.1:} the ERU is currently travelling to a target
\item \textbf{Fault 2.2:} the ERU is currently transporting
  a patient to a hospital 
\item \textbf{Fault 2.3:} the ERU is not currently on assignment
\end{itemize}
Recovery processes may differ for each of these scenarios, so we partition them (they can be amalgamated later if we discover that detection and recovery do not differ).  For some of these scenarios, recovery simply consists of providing the medical crew with transport and removing or repairing the original vehicle.  However, if the ERU is travelling to or transporting a patient, then in addition to these steps an alternative ERU must be directed to take over patient transportation, because waiting for repair or a replacement vehicle introduces unacceptable delay.   Figure~\ref{fig:fault2-FAV} presents an FAV for Fault 2.1 (FAVs of Faults 2.2. and 2.3 omitted due to space limitations). The fault is depicted as activating during the ERU nominal process `ServiceRescue'.  

Initially for Fault 2 we assume that the ERU crew will be the first to detect the problem.  However in some situations the ERU
crew may be incapacitated and unable to communicate with the SoS (e.g., they are injured themselves, or the radio equipment is damaged).  In this case, detection falls to the call centre operator. The operator may be alerted to the problem, e.g., via phone calls from members of the public.  Or the operator may raise the alarm after being unable to contact the ERU within a reasonable timeframe.  For this reason, Figure~\ref{fig:fault2-FAV} presents two possible detection events, each resulting in a separate recovery process; one recovery process is initiated if the ERU themselves detect and report the fault, and the other recovery strategy is initiated by the call centre operator.  

\begin{figure}[!t]
	\centering
        \includegraphics[width=1\columnwidth]{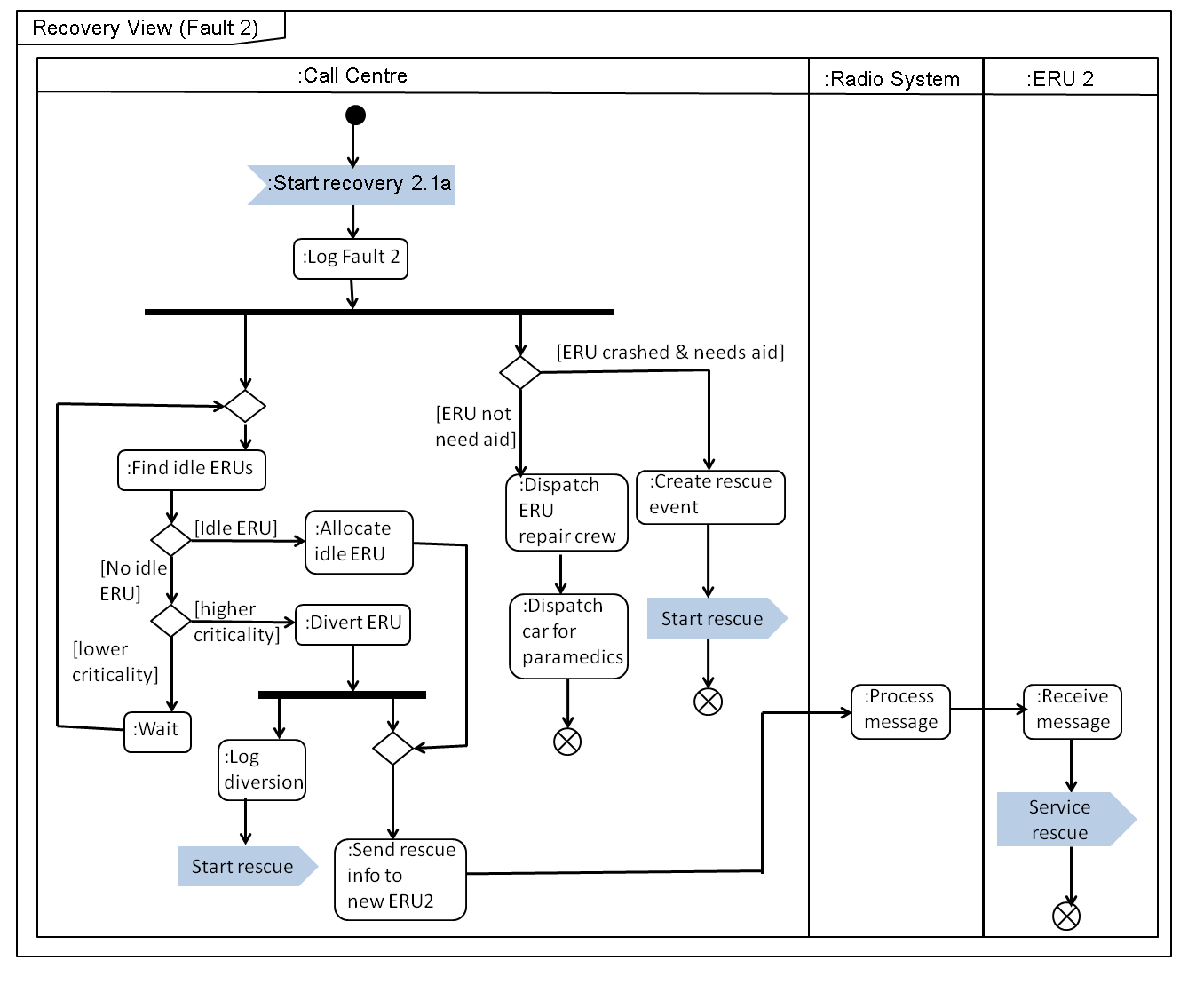}
	\caption{Fault Recovery View for Fault 2 (before reaching the target)}
	\label{fig:fault2-Recovery}
\end{figure}
Figure~\ref{fig:fault2-Recovery} presents a recovery view for Recovery Process 2.1a.  This recovery is appropriate for Fault 2.1, and is initiated by the call centre (as shown in Figure~\ref{fig:fault2-FAV}).  An FMAF recovery view shows what action is taken by each CS as the SoS recovers from the fault.  In this case, Fault 2 was activated before the ERU reached the target, so recovery involves finding a new ERU, (`ERU 2'), to attend the original target, (possibly diverting an ERU on a lower priority assignment). In addition, the original vehicle and its crew must also be dealt with, which may involve dispatching repair teams, or initiating a new rescue event if the ERU crew themselves require medical aid.   

As noted earlier, either the ERU or the call centre can detect Fault 2.1 and initiate recovery.   The recovery strategies employed by the two CSs are not the same.  If the ERU is able to `detect' the fault on behalf of the SoS, then their first action will be to contact the call centre operator via radio to report the problem and request appropriate aid for themselves and a replacement to take over their assignment.  However, if the ERU are unable to detect the fault for the SoS, then the responsibility falls to the call centre operator. This is the recovery process shown in Figure~\ref{fig:fault2-Recovery}.  The call centre operator may have received a report from a third party with details of the problem, or alternatively s/he may suspect that there is a problem, but not know what has happened or therefore what aid is required.  Figure~\ref{fig:fault2-Recovery} shows two parallel activities for the call centre operator during recovery: locating another ERU to take over the original assignment; and taking some actions to remedy the immobilised original vehicle and crew.  There is a conditional branch here: if the operator knows that the original ERU is simply broken down, the required actions include dispatching transportation for the crew and dispatching mechanics for the vehicle.  Alternatively, if the operator knows that the ERU has crashed, actions involve creating a new rescue event and following the usual procedure to dispatch medical aid to an emergency.  

The question still arises of what the operator should do if they suspect there is a problem (e.g., because the ERU has been uncontactable for some time) but does not have details of the nature of the problem (e.g., it's unknown whether there is a break down or an accident).  One possible strategy is to assume the worse case scenario immediately: the call centre operator could dispatch a vehicle with aid if the ERU is uncontactable for some predetermined period of time.  This will lead to crashed and broken down ERUs being recovered much more quickly than otherwise, but it risks wasting resources if - for example - the ERU is in a radio-shadowed area and not suffering any vehicle problems.  Alternatively, the ERU suppliers can elect to take some other action - e.g., installing devices that can self-report some types of vehicle problems; this requires a business case to consider the probable cost benefits to each CS affected by the activation of the fault and initiation of recovery.  There is a need to model and agree upon the correct recovery procedure in advance, because in this case the behaviour and installed devices of one CS (the ERU) need to be well understood by another CS (the call centre) in order to diagnose a problem; and because the call centre will need to take responsibility for initiating recovery action (with an associated cost) based on the expected behaviour of the ERU crew and/or hardware.  The FMAF models can be useful here for identifying different activities that can be affected by a fault (thereby aiding with partitioning faults which have similar causes but different recoveries, as is the case for Faults 2.1, 2.2. and 2.3) and clearly showing which CS can detect the problem and will be initiating recovery.

The FAV is also helpful for identifying where nominal procedures may need to be
modified as a result of designing recovery procedures. For example, after considering options for recovering from Fault 2.1, the reliability engineer may decide that the call centre should monitor the ERU's response times.  There is a concept of a timer in the call centre's nominal behaviour, visible in~\cite{Andrews&13d}; the reliability engineers can return to this nominal behaviour model and modify or extend is as necessary to permit the appropriate recovery.   

\section{Conclusions and future work}\label{sec:concs}
In this paper we have briefly demonstrated the use of the
COMPASS-developed FMAF approach for modelling fault tolerance in an
SoS at the architectural level.  We have built further on a case study
which was previously published with a single fault modelled; our
contribution here goes further by demonstrating the modelling of two
quite different faults in the same SoS. 

In this paper we present two faults which may be detected by more than
one CS.  For example, Fault 2.1 (ERU crashes or breaks down en route
to target) may be detected by the ERU crew, or alternatively by the
call centre.  Fault 3.2 (call centre transmits incorrect location)
presents an example of a fault which may be detected by two different
CSs, and also where the fault activation and detection/recovery may be
allocated to different CSs.  The identity of the detector is significant, because this dictates which recovery processes are needed.  For example, for Fault 2, in cases where the ERU detects the fault, details on ERU status and needs are available and narrow down the necessary recovery actions.  However, if the call centre acts as the detector and the ERU is uncontactable, the call centre does not necessarily have access to these details and must make some assumptions.    


We have employed two different methods of presenting the multiple detector roles in our FAV diagrams for Faults 2 and 3.  In Figure~\ref{fig:fault2-FAV} we have employed an interruptible region for the detection of Fault 2, and initiation of recovery by the ERU, and in the same diagram a separate interruptible region for the detection of Fault 2 and initiation of recovery by the call centre.  However, in Figure~\ref{fig:3b_FAV} we have employed a single interruptible region to depict detection of Fault 3 and initiation of recovery by either the ERU or the call centre.  There are advantages and disadvantages for each method.  Employing separate interruptible regions for the two different `dectector' roles (as in Figure~\ref{fig:fault2-FAV}) allows us to demonstrate clearly where the responsibility for the detection and recovery lies.  This is particularly pertinent for an SoS, where the separate constituent systems are independent and autonomous.  Representing the initiation of similar recovery processes together (e.g., Figure~\ref{fig:3b_FAV}), is useful for situations where separate recoveries need to be considered together, and ensures that the `flow' of processes is easily followed.  This is important for an SoS, where there may be events arising in different constituent systems in parallel.  The FMAF modelling approach is a useful
tool for reliability engineers, who may need to initiate some
negotiations and agreements between independently managed and funded
CSs, particularly in cases where one CS may need to accept
responsibility for detecting and recovering from faults introduced by
others.

There are several areas for future and ongoing work in this area of SoS fault tolerance, including:
modelling remaining faults in the TMS case study; including additional concepts from the dependability taxonomy~\cite{Avizienis04}; 
incorporating the newly identified extensions into the FMAF framework definition; and
the development of tools and techniques to
support linking between architectural models and fault analysis tools
(such as HiP-HOPS\footnote{http://hip-hops.eu}) and formal
verification techniques.


\section*{Acknowledgments}
The work presented here is supported by the EU Framework 7 Integrated
Project `Comprehensive Modelling for Advanced Systems of Systems'
(COMPASS, Grant Agreement 287829). For more information see
\url{http://www.compass-research.eu}.  The authors are grateful to
Insiel for their input on developing the case study.
	
\bibliographystyle{IEEEtran}
\bibliography{edsos}

\end{document}